\documentclass[12pt]{iopart}
\usepackage{graphicx}

\begin{document}

\title[]{Deterministic entanglement generation between a pair of atoms on different Rydberg states via chirped adiabatic passage}

\author{Jing Qian$^{\dagger,1,3}$ and Weiping Zhang$^{2,3}$}

\address{$^1$Department of Physics, School of Physics and Material Science, East China
Normal University, Shanghai 200062, People's Republic of China}
\address{$^2$Department of Physics and Astronomy, Shanghai JiaoTong University, Shanghai 200240, People's Republic of China}
\address{$^3$Collaborative Innovation Center of Extreme Optics, Shanxi University, Taiyuan, Shanxi 030006, People's Republic of China}
\ead{jqian1982@gmail.com}
\vspace{10pt}
\begin{indented}
\item[]November 2016
\end{indented}

\begin{abstract}
We develop a scheme for deterministic generation of an entangled state between two atoms on different Rydberg states via a chirped adiabatic passage, which directly connects the initial ground and target entangled states and also does not request the normally needed blockade effect.
The occupancy of intermediate states suffers from a strong reduction via two pulses with proper time-dependent detunings and the electromagnetically induced transparency condition. By solving the analytical expressions of eigenvalues and eigenstates of a two-atom system, we investigate the optimal parameters for guaranteeing the adiabatic condition. We present a detailed study for the effect of pulse duration, changing rate, different Rydberg interactions on the fidelity of the prepared entangled state with experimentally feasible parameters, which reveals a good agreement between the analytic and full numerical results.
\end{abstract}

%
%
\submitto{\JPB}

\section{Introduction}

Entanglement, as one of the unique quantum features, is not only an important resource for quantum information processing, quantum computation, and quantum metrology, but also a subject of great theoretical interests to understand the quantum physics. It has been demonstrated in a great variety of quantum systems such as photon pairs \cite{Stevenson06,Dousse10}, trapped ions \cite{Blatt08}, atomic ensembles \cite{Julsgaard01,Chou05} and nitrogen-vacancy centers \cite{Dutt07,Song15}. Owing to the weak interactions, the creation of entanglement between neutral atoms in the ground state has to resort to the additional enhancement approaches, such as using a high-$Q$ cavity to mediate the interaction between transient atoms \cite{Hagley97} and controlling interatomic collisions by the optical lattice \cite{Mandel03}. 

Recently, rapid developments of researches in Rydberg excitations provide a new route to the creation of entanglement with neutral atoms. The dipolar interaction between alkali-metal atoms in highly-excited Rydberg states becomes orders of magnitude stronger than the interaction between ground states, which inhibits the simultaneous excitation of two or more atoms into the same Rydberg state by the so-called blockade effect, entangling two atoms in the ground state or the Rydberg state \cite{Gallagher08,Comparat10}. The blockade mechanism in Rydberg atoms opens many new possibilities to exploit neutral atoms for the study of quantum computation and simulation \cite{Saffman10}. 
After the first proposal suggested by Jaksch and coworkers to implement a fast two-qubit entangling gate \cite{Jaksch00}, there has been a variety of theoretical and experimental works for realizing entangled state with Rydberg atoms by strong blockade, where the Rydberg interaction strength is large compared to the Rabi frequency of driving laser \cite{Unanyan02,Wilk10,Zhang10,Zuo10,Wuster13,Mobius13,Ebert15,Maller15,Garttner15}. 
For example, combing dissipation and Rydberg blockade can create various complex entangled states \cite{Rao13,Carr13,Rao14, Su15}. Especially, in a multi-level atomic system, the STImulated Raman Adiabatic Passage (STIRAP) combined with Rydberg blockade is manifest as an efficient way to resist spontaneous emission in producing entangled states \cite{Moller08,Yan11,Cano14,Idlas16}. The essence of STIRAP relies on the use of counterintuitive laser pulse sequence to transfer population by adiabatically following a dark state, which does not involve intermediate states \cite{Bergmann98,Petrosyan13}. More recently, an alternative scheme by using double adiabatic passage across a F\"{o}rster resonance for the implementation of two-qubit quantum gate without blockade is also proposed \cite{Beterov16}.

In the current work we develop a scheme for deterministically preparing an entangled state between two atoms on different Rydberg states (i.e. $(\left\vert sr\right\rangle+\left\vert rs\right\rangle)/\sqrt{2}$, $\left\vert s\right\rangle$ and $\left\vert r\right\rangle$ are Rydberg states) by a frequency-chirped adiabatic passage and without the requirement of strong blockade. We consider a pair of three-level ladder-type atoms that is initially prepared in the ground state and investigate its adiabatic transfer towards the target entangled state by using proper time-dependent detuning pulses \cite{Beterov11, Kuznetsova14,Qian15}. Such a level configuration with two or more Rydberg states contains richer nonlinear phenomena due to its complicated interactions between different Rydberg states \cite{Saffman09,Li14,Petrosyan14,Tian15}. According to the analytical expression of adiabatic eigenstate given by the perturbations, the temporal profile of pulses can be directly obtained, for achieving a high fidelity between the real final state and the target entangled state. The influence of Rydberg interactions, especially the exchange interaction between different Rydberg states, on the fidelity is also investigated. Our scheme can serve as a new avenue to the generation of robust and clean maximal entanglement between Rydberg atoms. Comparing to the common entangled states with one Rydberg state and one ground state, the obtained entangled state involving two different Rydberg states has more widely applications in the quantum technology.

\section{The two-atom Scheme}

\begin{figure}[ptb]
\centering
\includegraphics[width=4.2in,height=2.0in]{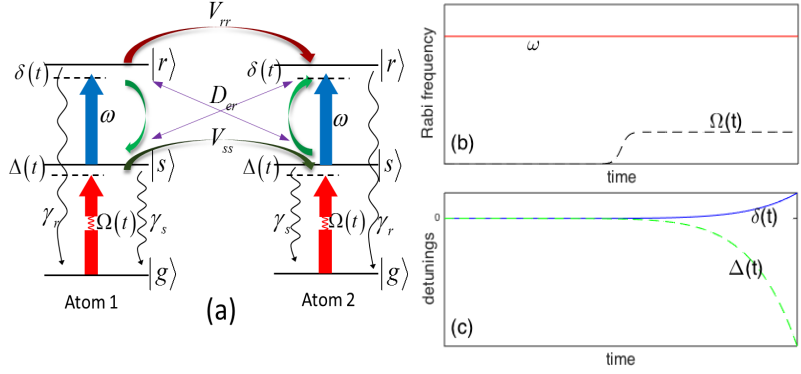}
\caption{(color online). (a) Schematic representation of the energy level structure for a pair of Rydberg atoms. (b)-(c) Time dependence of the required laser Rabi frequency $\Omega(t)$ (black dashed), microwave Rabi frequency $\omega$ (red solid), and the detunings $\delta(t)$ (blue solid) and $\Delta(t)$ (green dashed).}
\label{model}
\end{figure}

The physical setup for the entanglement generation, see Fig. \ref{model}(a), involves a pair of cold atoms with two highly-excited Rydberg states $\left\vert s\right\rangle$, $\left\vert r\right\rangle$ and one ground state $\left\vert g\right\rangle$. $\left\vert g\right\rangle$ and $\left\vert s\right\rangle$ are coupled with an effective laser Rabi frequency $\Omega$ and detuning $\Delta$ by an one- \cite{Hankin14} or two-photon \cite{Ye13} process. The Rydberg state $\left\vert s\right\rangle$ is further coupled to another adjacent Rydberg state $\left\vert r\right\rangle$ under microwave Rabi frequency $\omega$ and detuning $\delta$. 
Spontaneous decays from $\left\vert s\right\rangle$ (or $\left\vert r\right\rangle$) to $\left\vert g\right\rangle$ is denoted by $\gamma_{s}$ (or $\gamma_{r}$).

The Hamiltonian describing the system can be written as ($\hbar=1$ everywhere)
\begin{equation}
\mathcal{H} = \sum_{j=1,2}\mathcal{H}_{j}+\mathcal{V}_{ss}+\mathcal{V}_{rr}+\mathcal{D}_{sr},
\label{ham}
\end{equation}
where the single atom Hamiltonian is $\mathcal{H}_{j}=\Delta \sigma_{ss}^{j}+\delta \sigma_{rr}^{j}+\Omega(\sigma_{gs}^{j}+\sigma_{sg}^{j})+\omega(\sigma_{sr}^{j}+\sigma_{rs}^{j})$ with the atomic operators $\sigma_{\alpha\beta}^j=\left\vert \alpha\right\rangle\left\langle \beta\right\vert_{j}$ for atom $j$. The relevant interactions between Rydberg states are classified as the van der Waals (vdWs) \cite{Beguin13,Thaicharoen15} and the dipole-dipole (DD) interactions \cite{Li05,Li06}. If two atoms occupy same state $\left\vert s\right\rangle$ or $\left\vert r\right\rangle$, the intrastate interactions are described by the vdWs type, which are
\begin{eqnarray}
\mathcal{V}_{ss}&=\mathcal{V}_{0,ss}\sigma_{ss}^{1}\otimes\sigma_{ss}^{2},\\
\mathcal{V}_{rr}&=\mathcal{V}_{0,rr}\sigma_{rr}^{1}\otimes\sigma_{rr}^{2},
\end{eqnarray}
and if they occupy differently, the interstate exchange interaction described by the DD type is
\begin{equation}
\mathcal{D}_{sr}=\mathcal{D}_{0,sr}(\sigma_{sr}^{1}\otimes \sigma_{rs}^{2}+\sigma_{rs}^{1}\otimes \sigma_{sr}^{2}).
\end{equation}

Here, $\mathcal{V}_{0,ss}=C_{6}^{ss}/r^{6}$, $\mathcal{V}_{0,rr}=C_{6}^{rr}/r^{6}$ and $\mathcal{D}_{0,sr}=C_{3}^{sr}/r^{3}$, with $C_{6}^{ss(rr)}$ and $C_{3}^{sr}$ the dispersion coefficients and $r$ the interatomic distance. 
Note that the exchange interaction $\mathcal{D}_{0,sr}$ is predominantly responsible for preparing entangled pair state (exciton state) in F\"{o}rster resonance \cite{Ates08,Bettelli13,Schempp15,Schonleber15}.


There are various approaches for generating entanglement with Rydberg atoms. One typical way is using blockade effect in which the intrastate interaction is larger than the effective Rabi frequency between the ground and Rydberg states, giving rise to a singly-excited collective state (entangled state) $\left\vert E\right\rangle=1/\sqrt{2}(\left\vert gr\right\rangle+\left\vert rg\right\rangle)$ of two atoms \cite{Gaetan09,Labuhn14}. State $\left\vert E\right\rangle$ can further be extended to generate a many-atom entangled state $\left\vert E_{N}\right\rangle=\frac{1}{\sqrt{N}}\sum_{j=1}^{N}\left\vert g_1...r_j...g_N\right\rangle$ with one excitation shared by $N$ atoms in the ensemble \cite{Lukin01,Stanojevic09}.  Another way is relying on a STIRAP that can transfer the population onto an entangled state of two lower states $\left\vert D_{\infty}\right\rangle=1/\sqrt{2}(\left\vert ss\right\rangle-\left\vert gg\right\rangle)$ (subscript $\infty$ means $t\to\infty$), by adiabatically following a dark state \cite{Moller08}. This entangled state is robustly created irrespective of the precise values of couplings and interactions.

More recently, Rost's group proposes a scheme of creating entangled atom pairs via interstate resonant DD interactions between two different Rydberg states \cite{Wuster13}. After preparing twin atom clouds in a blockaded condition by strong vdWs interactions, they find the resonant DD interactions can give rise to a conversion from the blockaded state to a many-atom repulsive exciton state. The key for realizing this entangled exciton state is making a pair of atom clouds that only supports one collective excitation per cloud, which is the basic effect of strong blockade. As for one atom per cloud, by separately controlling the excitation of atom in each cloud, the two atoms can finally be prepared onto an entangled exciton state $\left\vert Ry\right\rangle =1/\sqrt{2}(\left\vert sr\right\rangle+\left\vert rs\right\rangle)$. A more general case where the entanglement is created between two distant atom clouds by using Rydberg dressing is predicted in \cite{Mobius13} and the excitation transport of exciton is also studied in \cite{Mobius11}.

Inspired by these works, we focus on preparing a high-fidelity entangled state $\left\vert Ry\right\rangle$ between two Rydberg states $\left\vert s\right\rangle$, $\left\vert r\right\rangle$ via a chirped adiabatic passage. Comparing to $\left\vert E\right\rangle$ and $\left\vert D_{\infty}\right\rangle$, the production $\left\vert Ry\right\rangle$ containing two Rydberg states is highly stable and easy to be manipulated with external fields. Differing from Rost's works, the time-dependence of the fields (lasers, detunings) is intuitively obtained from the presence of an adiabatic eigenstate, which directly connects the initial (ground) and target (entangled) states with a big suppression to the population of intermediate states. Also, the requirement for strong interactions is weak in our scheme.

\section{Chirped adiabatic passage}

For two atoms the evolution of system preserves the symmetry under the exchange of atoms so it is sufficient to rewrite the Hamiltonian (\ref{ham}) in the symmetric two-atom basis $\left\vert j\right\rangle=\{\left\vert gg\right\rangle,(\left\vert gs\right\rangle+\left\vert sg\right\rangle)/\sqrt{2},\left\vert ss\right\rangle,(\left\vert gr\right\rangle+\left\vert rg\right\rangle)/\sqrt{2},(\left\vert sr\right\rangle+\left\vert rs\right\rangle)/\sqrt{2},\left\vert rr\right\rangle\}$ ($j=1,2,...,6$), with $\left\vert 5 \right\rangle$ (=$\left\vert Ry \right\rangle$) the target entangled state. $\mathcal{H}$ can be given in a matrix form, as
\begin{equation}
\mathcal{H}=\left(
\begin{array}
[c]{cccccc}%
0 & \sqrt{2}\Omega & 0 & 0 & 0 & 0\\
\sqrt{2}\Omega & \Delta & \sqrt{2}\Omega & \omega & 0 & 0\\
0 & \sqrt{2}\Omega & S & 0 & \sqrt{2}\omega & 0\\
0 & \omega & 0 & \delta & \Omega & 0\\
0 & 0 & \sqrt{2}\omega & \Omega & D & \sqrt{2}\omega\\
0 & 0 & 0 & 0 & \sqrt{2}\omega & R
\end{array}
\right),  
\label{Hamtwo}%
\end{equation}
where the effective detunings are $S=2\Delta+\mathcal{V}_{0,ss}$, $R=2\delta+\mathcal{V}_{0,rr}$ and $D=\Delta+\delta+\mathcal{D}_{0,sr}$. 

We consider a two-atom eigenstate in a generalized form, which is
\begin{equation}
\left\vert \lambda(t)\right\rangle=a_{1}\left\vert 1\right\rangle+a_{2}\left\vert 2\right\rangle+a_{3}\left\vert 3\right\rangle+a_{4}\left\vert 4\right\rangle+a_{5}\left\vert 5\right\rangle+a_{6}\left\vert 6\right\rangle,
\label{Bt}
\end{equation}
with $|a_{j}(t)|^2$ standing for the population in each basis $\left\vert j\right\rangle$. Our target is allowing $\left\vert\lambda(0)\right\rangle=\left\vert 1\right\rangle$ and $\left\vert\lambda(\infty)\right\rangle=\left\vert 5\right\rangle$, and the population of intermediate states is deeply suppressed, satisfying $|a_{1}(t)|^2+|a_{5}(t)|^2\gg|a_{2}(t)|^2+|a_{3}(t)|^2+|a_{4}(t)|^2+|a_{6}(t)|^2$. 
For that purpose, due to the adiabatic theorem of quantum mechanics, a novel non-degenerate adiabatic eigenstate $\left\vert \lambda_0(t)\right\rangle$ is required in which the system can evolve without making any transitions, when the level spacing of $\left\vert \lambda_0(t)\right\rangle$ with other eigenstates is much larger compared to the changing rate of the Hamiltonian $\mathcal{H}(t)$ \cite{Pu07}.
Commonly speaking, this target could be realized by preparing a dark state, which is fully unaffected by the short lifetime of intermediate excited state, as similar as in a three-state system due to quantum interference \cite{Vitanov16}. However, accounting for the complex energy levels and interactions, a quasi-dark adiabatic eigenstate $\left\vert \lambda_0(t)\right\rangle$ is obtained instead, in which the population of intermediate states can not be perfectly suppressed.

Direct solving the secular equation of Hamiltonian (\ref{Hamtwo}) gives rise to six sets of eigenvalues $\lambda_{k}(t)$ and eigenstates $\left\vert \lambda_k(t)\right\rangle$ ($k=0,1,2,...,5$). To extract the adiabatic eigenstate $\left\vert \lambda_{0}\right\rangle$ among them, we first utilize the perturbation method in the limit of $\Omega/\omega\ll 1$ to simplify the secular equation with eigenvalues $\lambda_k$. The resulting smallest eigenvalue to the third order of $\Omega/\omega$ can be given by
\begin{equation}
\lambda_{0}^{a}\approx\frac{2\delta\Omega^{2}}{\omega^{2}-\delta\Delta}.
\label{zeroener}
\end{equation}

Besides, the nearest and next-to-nearest eigenvalues with respect to $\lambda^a_0$ are given by
\begin{eqnarray}
\lambda_{1}^{a}&\approx\frac{2\omega^2(R+S)-DRS}{4\omega^2-DR-DS-RS},\\
\lambda_{2}^{a}&\approx\frac{(\Delta+\delta)+\sqrt{(\Delta+\delta)^2-4(\Delta\delta-\omega^2)}}{2},
\end{eqnarray}
where the superscript $a$ stands for analytical results. The condition of $\Omega/\omega\ll 1$ indicates that the microwave driving (labeled by $\omega$) is stronger than the laser driving (labeled by $\Omega$), i.e. the electromagnetically induced transparency (EIT) condition \cite{Garttner14}.
From Eq. (\ref{zeroener}) it implies $\delta\Delta<0$ is necessary (avoiding the divergence of $\lambda_0^a$) for keeping $|\lambda_0^a|$ a small value, which plays a crucial role for the existence of such a quasi-dark state $\left\vert \lambda_{0}\right\rangle$ for realizing an adiabatic evolution onto the entangled state $\left\vert 5\right\rangle$. 

Accordingly, the population in this adiabatic eigenstate $\left\vert\lambda_0^a(t)\right\rangle$ is approximately described as
\begin{eqnarray}
\label{a2}%
a_{2}^{a}  & \approx\frac{\sqrt{2}\delta\Omega}{\omega^{2}-\delta\Delta}a_{1}^a,\\
\label{a3}
a_{3}^{a}  &  \approx-\frac{2\delta}{S+2\delta}a_{1}^a,\\
\label{a4}
a_{4}^{a}  & \approx-\frac{\sqrt{2}S\Omega}{\left(  S+2\delta\right)  \omega}a_{1}^a,\\
\label{a5}
a_{5}^{a}  &  \approx\frac{\sqrt{2}S\delta}{\left(  S+2\delta\right)  \omega}a_{1}^a,\\
\label{a6}
a_{6}^{a}  &  \approx-\frac{2S\delta}{\left(  S+2\delta\right)  R}a_{1}^a,%
\end{eqnarray}
with $a_1^a$ solved from the conservation: $|a_1^a|^2+|a_2^a|^2+|a_3^a|^2+|a_4^a|^2+|a_5^a|^2+|a_6^a|^2=1$, taking form of
\begin{equation}
|a_{1}^{a}|   \approx\frac{1}{\sqrt{  1+\frac{2\delta^{2}\Omega^{2}}{\left(  \omega
^{2}-\delta\Delta\right)  ^{2}}+\frac{4\delta^{2}\omega^{2}\left(  R^{2}%
+S^{2}\right)  +2R^{2}S^{2}\left(  \Omega^{2}+\delta^{2}\right)  }{\left(
S+2\delta\right)  ^{2}\omega^{2}R^{2}}} }.
\label{a1v}
\end{equation}

Initially, $\delta=0$ and $\Omega/\omega\ll 1$ ensure the preparation of $\left\vert\lambda_{0}^{a}(0)\right\rangle=\left\vert 1\right\rangle$ ($|a^a_1|^2=1$). According to Eq. (\ref{a5}), we see $|a_5^a|$ increases with $S$ or $\delta$. For the purpose of letting $\left\vert\lambda_{0}^{a}(\infty)\right\rangle=\left\vert 5\right\rangle$ at the final time, one possible way is adiabatically tuning $|\Delta(t)|$ and $|\delta(t)|$ from 0 to large positive values, during which its instantaneous eigenstate $\left\vert\lambda_{0}^{a}(t)\right\rangle$ persists. 
Based on the above analysis, the time dependence of the envelope for the chirped detunings can be intuitively obtained, as presented in Fig. \ref{model}(c) where $|\delta(t)|$ and $|\Delta(t)|$ are both adjusted from zero to positive values with opposite signs for $\delta\Delta<0$. The temporal profiles of laser and microwave pulses are also displayed in Fig. \ref{model}(b), where $\omega$ is fixed and $\Omega(t)$ changes slightly with time for keeping the adiabaticity. Note that $\omega\gg\Omega_{max}$ (the peak value of $\Omega(t)$) is always kept. Comparing to \cite{Wuster13} where the required shapes of microwave and detuning pulses are relatively complicated, the pulses needed in our scheme are more flexible that only should satisfy the adiabatic evolution of $\left\vert\lambda_{0}^{a}(t)\right\rangle$. Based on the analytical expressions of eigenvalues $\left\vert\lambda_{0}^{a}(t)\right\rangle$ and $\left\vert\lambda_{1}^{a}(t)\right\rangle$ [see Eqs. (7)-(8)], the adiabatic condition is given by \cite{Bohm51}
\begin{equation}
\max\{|d\delta(t)/dt|,|d\Delta(t)/dt|\}\ll |\lambda_{0}-\lambda_{1}|^2,
\label{ac}
\end{equation}
with $\lambda_0$ and $\lambda_1$ the eigenvalues of the target adiabatic and its nearest neighboring non-adiabatic eigenstates.

In addition, from Eqs. (\ref{a3}) and (\ref{a6}), we notice that if $\mathcal{V}_{0,ss}$ or $\mathcal{V}_{0,rr}$ is compensated by proper detunings, i.e. $S$ or $R$ vanishes, the emerging anti-blockade effect would enhance the population in the doubly excited state $\left\vert ss\right\rangle$ or $\left\vert rr\right\rangle$, rather than the target entangled state $(\left\vert sr\right\rangle+\left\vert rs\right\rangle)/\sqrt{2}$. To avoid this, a direct way is allowing $\Delta(\delta)$ and $\mathcal{V}_{0,ss}(\mathcal{V}_{0,rr})$ to have same signs. In experiment, it is feasible to prepare atoms in Rydberg state $\left\vert s\right\rangle$ with attractive intrastate interactions $\mathcal{V}_{0,ss}<0$ \cite{Pritchard10}.

\section{Fidelity of adiabatic passage} \label{ffde}

The realistic dynamics of system is governed by the two-atom master equation
\begin{equation}
\frac{d\rho(t)}{dt}=-i[\mathcal{H},\rho(t)]+\sum_{j=1,2}\mathcal{L}_{j}[\rho]
\label{master}
\end{equation}
with the Lindblad operators $\mathcal{L}_{j}[\rho]$
\begin{equation}
\mathcal{L}_{j}= \gamma_s(\sigma_{gs}^j\rho\sigma_{sg}^j-\frac{\{ \sigma_{ss}^j, \rho \}}{2})
+\gamma_r(\sigma_{gr}^j\rho\sigma_{rg}^j-\frac{\{ \sigma_{rr}^j, \rho \}}{2})
\label{Lind}
\end{equation}
where $\gamma_{s(r)}$ is phenomenologically introduced to present the spontaneous decay of Rydberg state $\left\vert s\right\rangle$ or $\left\vert r\right\rangle$. $\rho(t)$ is the density matrix operator whose diagonalized elements $\rho_{jj}(t)$ stand for the real population dynamics of $\left\vert j\right\rangle$.

In the full numerical simulation, we consider a pair of chirped (time-dependent) detuning pulses, modeled by
\begin{equation}
\delta(t)  = \frac{\delta_{max}}{t_{max}^{m}}\times t^{m},\Delta(t)  = \frac{\Delta_{max}}{t_{max}^{m}}\times t^{m}
\label{detus}
\end{equation}
where $\delta_{max}$ and $\Delta_{max}$ are the peak values of $\delta(t)$ and $\Delta(t)$, and $m$ a variable that characterizes the changing rate. $t_{max}$ is the pulse duration. The laser pulse $\Omega(t)$ is modeled as a slowly-changing function, 
\begin{equation}
\Omega(t) = \frac{\Omega_{max}}{2}(1+\tanh(\frac{t-0.5t_{max}}{T_{\Omega}})),
\end{equation}
with the peak value $\Omega_{max}$ and the pulse width $T_{\Omega}$. 

To show the feasibility of scheme, experimentally realistic parameters are adopted in the numerical calculations. We choose $\Omega_{max}$=5.0MHz, $\omega$=20MHz ($\omega>\Omega_{max}$), $\delta_{max}=200$MHz, $\Delta_{max}=-1.0$GHz, $T_{\Omega}=1.0\mu$s. The intrastate vdWs interactions should be $\mathcal{V}_{0,ss}<0$ (same sign as $\Delta(t)$), $\mathcal{V}_{0,rr}>0$ (same sign as $\delta(t)$) as suggested. To be specific, we take $^{87}$Rb atoms in $n\approx57$, and $\left\vert s\right\rangle=\left\vert nP_{3/2},m_j=1/2\right\rangle$, $\left\vert r\right\rangle=\left\vert nD_{3/2},m_j=3/2\right\rangle$. 
Note that the vdWs coefficients for Rydberg states $\left\vert 57P_{3/2}\right\rangle$, $\left\vert 57D_{3/2}\right\rangle$ are both angular-dependent, so we focus on the case of $\theta=\pi/3$, where $\theta$ is defined by the angle between interatomic displacement and the quantization axis. Accounting for the results in Refs. \cite{Reinhard07} that the scaled energy shifts of $\left\vert s\right\rangle$ and $\left\vert r\right\rangle$ are approximately $-85$ and 100 respectively, it leads to the vdWs coefficients $C_{6}^{ss}\approx -2\pi\times152$GHz$\mu$m$^6$ and $C_{6}^{rr}\approx 2\pi\times230$GHz$\mu$m$^6$. For the DD exchange interaction between $\left\vert s\right\rangle$ and $\left\vert r\right\rangle$, we use the formula  $C_3^{sr}\approx \frac{3(3\sin^2\theta-2)}{32\pi\epsilon_0}n^{*4}$ \cite{Petrosyan14} that gives rise to $C_3^{sr}\approx2\pi\times200$MHz$\mu$m$^3$ when $\theta\approx\pi/3$. With these dispersion coefficients we could estimate the interaction strengths $\{\mathcal{V}_{0,ss},\mathcal{V}_{0,rr},\mathcal{D}_{0,sr}\}=\{-30,45,7.0\}$MHz for two atoms separated by a distance of 5.63$\mu$m \cite{Singer05,Walker08}. In addition, the effective lifetime of $\left\vert s\right\rangle$ and $\left\vert r\right\rangle$ are 450$\mu$s and 200$\mu$s at 0 K, giving to the spontaneous decays $\gamma_s=$2.0kHz, $\gamma_r=$5.0kHz \cite{Beterov09}.

With these parameters, we numerically solve the master equation (\ref{master}) and obtain the fidelity which takes the expression of \cite{Rao13}
\begin{equation}
F^{(a)}(t)=\left\langle \lambda^{(a)}_{0}(t) \right\vert \rho(t) \left\vert  \lambda^{(a)}_{0}(t)\right\rangle
\end{equation}
where $\left\vert\lambda^{(a)}_0\right\rangle$ is the numerical (analytical) adiabatic eigenstate and 
$\rho(t)$ is the real population dynamics solved from (\ref{master}). 
It is worth noting here that $\left\vert \lambda^{(a)}_{0}\right\rangle$ is of quasi-dark property that must be affected by the instantaneous population in the intermediate states during the transfer. Thus, keeping $t_{max}\ll \gamma^{-1}_{s},\gamma^{-1}_{r}$ is a possible way to reduce the loss from middle states \cite{Unanyan02}. Besides, we are particularly interested in the fidelity at $t=t_{max}$ (the end of pulse) which also stands for the final population in the entangled state $\left\vert 5\right\rangle$.

 \begin{figure}[ptb]
\includegraphics[width=3.5in,height=2.7in]{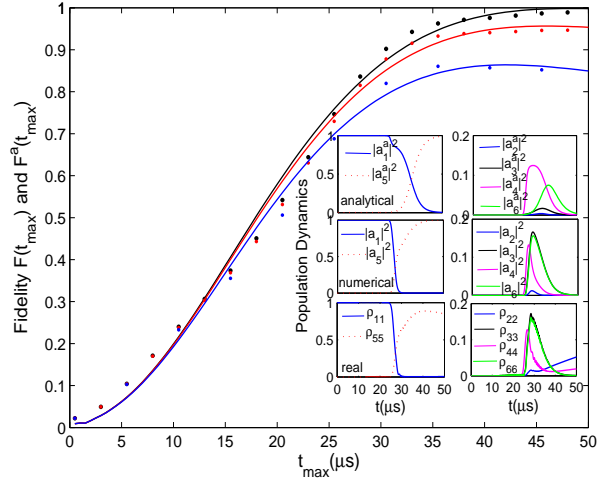}
\centering
\caption{(color online). Comparison of fidelities obtained by analytical ($F^a(t_{max})$, dotted curve) and numerical solutions ($F(t_{max})$, solid curve). From top to bottom, we use $\gamma_s$=$\gamma_r$=0.0 (black), $\gamma_s$=$\gamma_r$=1.0kHz (red), $\gamma_s$=2.0kHz and $\gamma_r$=5.0kHz (blue).  Insets show the time dependence of populations $|a^a_{j}(t)|^2$ (top panels), $|a_{j}(t)|^2$ (middle panels), $\rho_{jj}(t)$ (bottom panels) in the case of $\gamma_s$=2.0kHz, $\gamma_r$=5.0kHz and $t_{max}=50\mu$s. The changing rate $m=8$.}
\label{presult1}
\end{figure}

In Fig. \ref{presult1} we compare the fidelity $F(t_{max})$ (solid curve) solved from numerically diagonalizing Hamiltonian (\ref{Hamtwo}) with its analytical form $F^a(t_{max})$ (dotted curve). 
Both starting from the initial product state $\left\vert gg\right\rangle$,  a good agreement is found between $F(t_{max})$ and $F^a(t_{max})$ as $t_{max}$ is adjusted from 0 to 50$\mu$s. In addition, it shows that the fidelity reduces by less than 0.15 when $\gamma_{s(r)}$ increases from 0 (black solid and dotted), 1.0kHz (red solid and dotted) to a more realistic case with $\gamma_{s}$=2.0kHz, $\gamma_{r}$=5.0kHz (blue solid and dotted), since $\gamma_{s(r)} t_{max}\ll 1$ is always fulfilled. Furthermore, we also study a room temperature $T=300K$ with the decay rates $\gamma_{s}=7.0$kHz, $\gamma_{r}=10.0$kHz \cite{Beterov09}, and find the maximum of fidelity $F$ only attains 0.72 when $t_{max}\approx35\mu$s (not shown). 
Insets of Fig. \ref{presult1} presents a detailed description for the time-dependent populations $|a^a_{j}(t)|^2$ (top panels), $|a_{j}(t)|^2$ (middle panels) following the adiabatic eigenstate $\left\vert\lambda^{a}_{0}(t)\right\rangle$ and $\left\vert\lambda_{0}(t)\right\rangle$, as well as the real population dynamics $\rho_{jj}(t)$ (bottom panels) in the case of $\gamma_s$=2.0kHz, $\gamma_r$=5.0kHz. 
The perfect agreement between $|a_{j}(t)|^2$ and $\rho_{jj}(t)$ (except $\rho_{55}(\infty)$ is a bit small) confirms the existence of such a quasi-dark adiabatic eigenstate $\left\vert\lambda_{0}(t)\right\rangle$ that indeed can be adiabatically followed, resulting in an efficient population transfer from $|a_1|^2$($\rho_{11}$)$\to$$|a_5|^2$($\rho_{55}$). Note that the analytical solutions $|a^a_{j}(t)|^2$ have a slight deviation from the other two due to the perturbations.

The reason for poor fidelity ($F\approx0.862$) is mainly caused by the non-ignorable population in intermediate states, see the right column of insets. States $\left\vert 3,4,6\right\rangle$ are indeed occupied during the adiabatic transfer. If the decay rates are smaller than 1.0kHz the fidelity can attain more than 0.95. One way to improve the fidelity of our scheme is searching for suitable Rydberg states with longer lifetime. As in Rost's work \cite{Wuster13}, a higher fidelity is obtained in the timescale of a few $\mu$s with strong and short pulses; however it does not work here because we need to follow the evolution of a quasi-dark eigenstate that means a longer pulse duration will lead to better adiabaticity. So the decay rates of intermediate states must be an obstacle for the performance of the scheme.
In the following calculations, to focus on the effects of other parameters, we will ignore the decays by using $\gamma_s=\gamma_r=0$, $t_{max}=50\mu$s.

\begin{figure}[ptb]
\includegraphics[width=3.6in,height=2.7in]{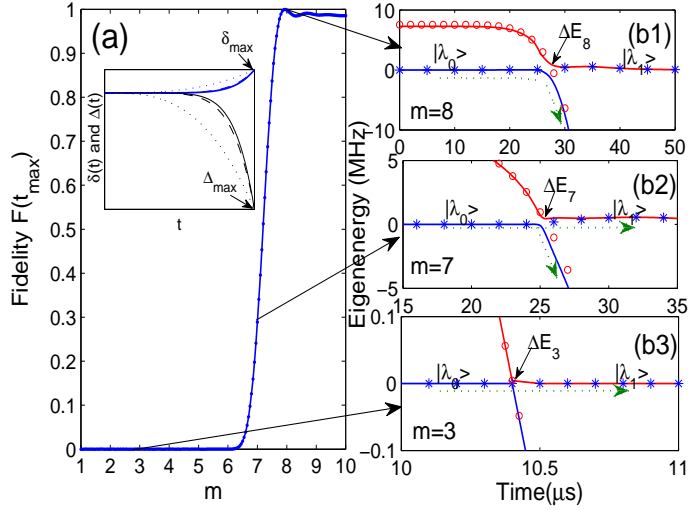}
\centering
\caption{(color online). (a) The fidelity with respect to the changing rate $m$ of the chirped detuning pulses. Inset: The variation of $\delta(t)$ and $\Delta(t)$ for different $m$ values, $m$=3.0 (dotted curve), 7.0 (dashed cruve), 8.0 (solid curve). (b1)-(b3) Amplified eigenenergies $\lambda_0$ and $\lambda_1$ near the avoided crossing for $m=(8.0,7.0,3.0)$. The analytical solutions $\lambda^a_0$ (blue stars) and $\lambda^a_1$ (red circles) are presented. The green dotted arrows denote the direction of real population transfer. }
\label{presult6}
\end{figure}

Next we investigate the effect of changing rate $m$ of chirped detuning pulses on the fidelity $F(t_{max})$. As displayed in Fig. \ref{presult6}a with $m$ increasing from 1.0 to 8.0, we see the fidelity sharply grows to be 1.0 after $m=6.0$. Through the definition of detunings (\ref{detus}), it is clear that $m$ value only determines the changing rate, rather than the initial and peak values that are fixed [see the inset of Fig.\ref{presult6}(a)]. The reason for that can be qualitatively understood from the eigenenergy spectrum of $\lambda_0(t)$ and $\lambda_1(t)$, labeled by solid curves as plotted in Fig.\ref{presult6}(b1)-(b3). The presence of an avoided crossing (near-degenerate energy levels) between the adiabatic eigenstate $\left\vert \lambda_0\right\rangle$ and its adjacent eigenstate $\left\vert \lambda_1\right\rangle$ is revealed. For comparison we also plot the analytical expressions of eigenvalues $\lambda^a_0$ (blue stars) and $\lambda^a_1$ (red circles) that are well consistent with the numerical results before the avoided crossing $t_{cro}$. But, it shows a reversal at $t>t_{cro}$, i.e.  $\lambda^a_0=\lambda_1$ and $\lambda^a_1=\lambda_0$. Because in the perturbation method, we have simply assumed $\left\vert\lambda^a_0\right\rangle$ is always the adiabatic eigenstate with its eigenvalue $|\lambda^a_0|$ smaller than other states during the whole adiabatic process. In a real process, $\left\vert\lambda_1\right\rangle$ instead of $\left\vert\lambda_0\right\rangle$ becomes the lowest energy state after the avoided crossing.

The location of the avoided crossing can be obtained exactly by considering $\lambda^a_0=\lambda^a_1$, leading to
\begin{equation}
t_{cro} \approx t_{max}(\frac{-\mathcal{V}_{0,rr}-\mathcal{V}_{0,ss}}{2(\delta_{max}+\Delta_{max})})^{\frac{1}{m}}.
\label{tcro}
\end{equation}

At $t=t_{cro}$, the addition of two detunings ($\Delta+\delta$) is compensated by the exchange energy $\mathcal{D}_{0,sr}$, leading to $D=0$. Hence, it is easy for us to solve the optimal $\mathcal{D}_{0,sr}$ value by considering $\mathcal{D}_{0,sr}=|\Delta(t_{cro})|-|\delta(t_{cro})|$. By using $\mathcal{V}_{0,rr}=45$MHz, $\mathcal{V}_{0,ss}=-30$MHz as suggested, we obtain $t_{cro}\approx 27.89\mu$s and the resulting $\mathcal{D}_{0,sr}$ is 7.5MHz, which is very close to the value 7.0MHz used in the numerical simulations.

Decreasing $m$ leads to the energy spacing $\Delta E_n$ at $t=t_{cro}$ rapidly reduces, that is $\Delta E_8$(2.232MHz)$>$$\Delta E_7$(0.934MHz)$>$$\Delta E_3$(0.0043MHz). A small energy spacing can easily give rise to the breakdown of the adiabaticity, see the right side of Eq. (\ref{ac}). Therefore, for $m=8$ the system evolutes along one isolated adiabatic eigenstate $\left\vert\lambda_0\right\rangle$ and enables the transfer of $\left\vert1\right\rangle\to\left\vert5\right\rangle$ (green dotted arrow). As for $m=7$ a partial number of population is transferred back into the ground state $\left\vert1\right\rangle$ following $\left\vert\lambda_1\right\rangle$ at $t>t_{cro}$. Turning to the case of $m=3$ with $\Delta E_3\ll \Delta E_8$, entire population persists settling on the ground state due to the energy degeneracy of $\left\vert\lambda_0\right\rangle$ and $\left\vert\lambda_1\right\rangle$ at $t=t_{cro}$. By the above analysis, we use $m=8.0$ in the following sections.

\section{Role of Rydberg interactions}

\begin{figure}[ptb]
\includegraphics[width=3.6in,height=2.2in]{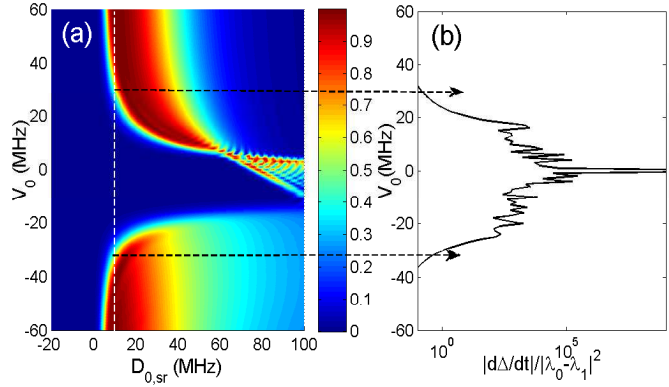}
\centering
\caption{(color online). (a) The fidelity of the generated entangled state as function of $\mathcal{D}_{0,sr}$ and $\mathcal{V}_0$; (b) A representation of the adiabaticity with $\mathcal{V}_0$ for $D_{0,sr}$=10.0 MHz.}
\label{presult4}
\end{figure}

To further investigate the criterion for high entanglement fidelity and understand the role of various Rydberg interactions, especially the interstate exchange interaction $\mathcal{D}_{0,sr}$, we study the fidelity $F(t_{max})$ in the space of $(\mathcal{D}_{0,sr},\mathcal{V}_0)$ as shown in Fig. \ref{presult4}(a), where we have assumed $\mathcal{V}_{0,rr}=\mathcal{V}_0$ and $\mathcal{V}_{0,ss}=-\mathcal{V}_0$ (opposite signs as required). 
Fig. \ref{presult4}(a) shows that, for a vanishing or negative $\mathcal{D}_{0,sr}$, the fidelity always vanishes irrespective of $\mathcal{V}_0$. 
The reason is by using $\Delta(t)+\delta(t)<0$, an appropriate and positive $\mathcal{D}_{0,sr}$ that can compensate for $\Delta(t)+\delta(t)$ is a key condition for realizing a high-fidelity entanglement.
Once $\mathcal{D}_{0,sr}$ is optimally determined, the creation of the maximally entangled state $\left\vert 5\right\rangle$ is robust, irrespective of the precise value of intrastate interactions $\mathcal{V}_0$. In other words, the strong blockade condition with the intrastate interaction $|\mathcal{V}_0|$ much larger than Rabi frequency $\Omega$ is not necessary. Our scheme has a good implementation in the partial blockade where the relevant parameters $|\mathcal{V}_0|$, $\Omega$, $\omega$ are comparable values \cite{Beguin13}.

Another interesting feature lies in that $F$ suddenly vanishes when $|\mathcal{V}_0|$ is too small, e.g. at $\mathcal{D}_{0,sr}=10$MHz and $|\mathcal{V}_0|<30$MHz, $F\approx 0$ as marked by a white dashed curve in Fig. \ref{presult4}(a). To understand this result, if $|\mathcal{V}_0|$ is too weak the doubly Rydberg state $\left\vert ss(rr)\right\rangle$ is highly detuned by the variation of $\Delta(\delta)$, the resulting final state will still be the initial ground state $\left\vert gg\right\rangle$. 
To be more clearly, we plot the adiabatic condition in Fig. \ref{presult4}(b) according to its definition (\ref{ac}), where the horizontal axis denotes the ratio of $|d\Delta(t)/dt|$ to the square of energy spacing between $ \lambda_1$ and $\lambda_0$, the vertical axis denotes $\mathcal{V}_0$. If $|d\Delta/dt|/|\lambda_0-\lambda_1|^2\gg 1$ it means the adiabaticity breaks. Actually, the ratio being much larger than one within the regime of $|\mathcal{V}_0|<30$ is observed, which exactly confirms the prediction that the adiabaticity really breaks there, leading to a poor fidelity.

\begin{figure}[ptb]
\includegraphics[width=3.6in,height=1.7in]{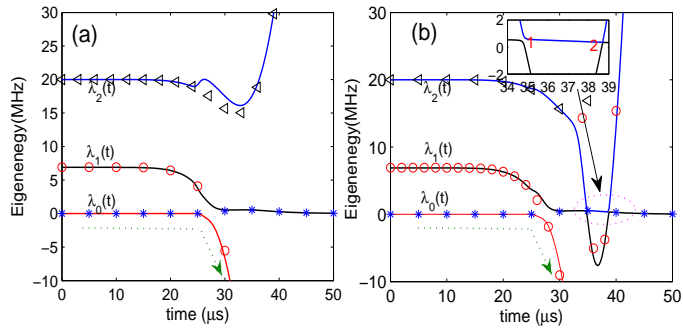}
\centering
\caption{(color online). Eigenenergy spectrum for the two-atom scheme. $\lambda_{0}(t)$ is eigenenergy for the target adiabatic eigenstate and $\lambda_{1}(t)$ ($\lambda_{2}(t)$) for the nearest (next-to-nearest) neighboring non-adiabatic eigenstate. (a) $V_0=60$MHz and (b) $V_0=-60$MHz. $\mathcal{D}_{0,sr}=10$MHz. Inset of (b): The amplification for two avoided crossings between $\lambda_{1}(t)$ and $\lambda_{2}(t)$ when $t\in(34,39)\mu$s. Analytical solutions of $\lambda^a_{0}(t)$, $\lambda^a_{1}(t)$, $\lambda^a_{2}(t)$ are respectively displayed by blue stars, red circles and black triangles. The green dotted arrows denote the real direction of population transfer. }
\label{presult5}
\end{figure}

To further understand the eigenenergy properties and compare the effect of intrastate Rydberg interactions $\mathcal{V}_{0}$ with different signs, we show the eigenvalues involving $\lambda_0$, $\lambda_1$ and $\lambda_2$ for $\mathcal{V}_{0}$=60MHz and -60MHz, with respect to Fig. \ref{presult5}(a) and (b).
In (a) one could obviously see that $\lambda_{0}(t)$ separates from $\lambda_{1}(t)$ and $\lambda_{2}(t)$ by large energy spacings. In particular, the smallest energy spacing between $\lambda_{0}(t)$ and $\lambda_{1}(t)$ at $t=t_{cro}$  is more than 2.5MHz (good adiabaticity). So state $\left\vert\lambda_{0}(t)\right\rangle$ is a good adiabatic eigenstate along which the full population transfer from $\left\vert 1\right\rangle\to\left\vert 5\right\rangle$ can carry out. The analytical solutions $\lambda^a_0$(blue stars), $\lambda^a_1$(red circles), $\lambda^a_2$(black triangles) perfectly agree with the numerical results except for the energy exchange of $\lambda^a_0$ and $\lambda^a_1$ after $t=t_{cro}$ caused by the assumption.

Whereas, in (b) there exists two extra avoided crossings between $\lambda_{1}(t)$ and $\lambda_{2}(t)$ [see the inset for an amplification], where the energy spacing tends to be zero. The presence of two extra avoided crossings is understandable where $\Delta(t)$ and $\delta(t)$ are respectively compensated by suitable $\mathcal{V}_{0,ss}$ and $\mathcal{V}_{0,rr}$. The resulting effective detunings $S$ and $R$ closing to zero may lead to the instantaneous occupancy on $\left\vert ss\right\rangle$ and $\left\vert rr\right\rangle$. Although $\left\vert\lambda_{1}(t)\right\rangle$ and $\left\vert\lambda_{2}(t)\right\rangle$ nearly degenerate in this case, we stress that the revival of the entangled state fidelity comes from the existence of an isolated adiabatic eigenstate $\left\vert\lambda_0\right\rangle$, which is luckily not influenced by these two avoided crossings. Therefore, as similar to (a) the high-fidelity entanglement generation remains in the case of $\mathcal{V}_0=-60$MHz. The reason for small deviations between numerical $\lambda_1$($\lambda^a_1$) and analytical results $\lambda_2$($\lambda^a_2$) in (b) comes from that the eigenenergy redistributes between $\left\vert\lambda_{1}(t)\right\rangle$ and $\left\vert\lambda_{2}(t)\right\rangle$ in the avoided crossing regimes.

\begin{figure}[ptb]
\includegraphics[width=2.4in,height=1.9in]{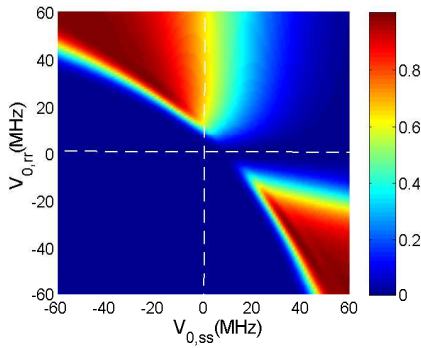}
\centering
\caption{(color online). The fidelity in $(\mathcal{V}_{0,ss},\mathcal{V}_{0,rr})$ space with $D_{0,sr}$=10 MHz. }
\label{presult7}
\end{figure}

Finally, we also show the fidelity with respect to different intrastate interactions $\mathcal{V}_{0,ss}$ and $\mathcal{V}_{0,rr}$, as displayed in Fig. \ref{presult7} which confirms that $\mathcal{V}_{0,ss}\mathcal{V}_{0,rr}<0$ is necessary for achieving the entanglement with high fidelity. Beyond that regime if $\mathcal{V}_{0,ss}\mathcal{V}_{0,rr}>0$ the fidelity is always very low no matter how to change them.

\section{Conclusion}

To conclude we have proposed a novel scheme for producing high-fidelity maximal entanglement between two different Rydberg states in a two-atom system based on a chirped adiabatic passage. Different from typical STIRAP technique that requires a pair of laser pulses performing in a counterintuitive order, two slowly-varying detuning pulses ($\Delta(t)$ and $\delta(t)$) with suitable envelopes enable an adiabatic connection between the initial ground and target entangled states. For realistic parameters used in the calculations we see that the initial population prepared in the ground state can efficiently evolve into a desired entangled state within tens of microseconds.
The scheme does not demand the interaction strengths to be in the strong blockade regime that is much larger than the laser Rabi frequency.
Once the temporal profiles of the detuning pulses and the atomic interactions, especially the exchange interaction between two atoms in different Rydberg states, are moderate and optimized, the entanglement which is insensitive to the precise value of intrastate interactions between same Rydberg energy levels, will be robustly created. Our method may provide a new route to the entanglement between two atoms on different Rydberg states with high fidelity.

In comparison with other methods using dissipation for the creation of an entangled steady state with high-fidelity $\sim0.99$ \cite{Carr13,Rao13,Rao14}, here, the obtained fidelity of entangled state is slightly low $\sim0.862$ with realistic parameters $\gamma_{s}$=2.0kHz, $\gamma_{r}$=5.0kHz and $t_{max}$=50$\mu$s. The main loss of fidelity for the entanglement generation is caused by the spontaneous decay from population in intermediate states during the adiabatic process, which is impossible to be fully suppressed due to the reliance of a quasi-dark adiabatic eigenstate. Also, as comparing to Rost's works, our scheme provides a flexible way to optimize time-dependent pulses based on the adiabatic evolution of a quasi-dark eigenstate. 
In the future, we will further explore the ways to improve the fidelity of entanglement with more appropriate Rydberg states, e.g. involving longer lifetimes or larger Rabi frequencies, or by using other approaches such as shortcut to adiabaticity \cite{Chen10}, adiabatic rapid passage \cite{Malinovsky01} and so on. Besides, we will extend our model to a $N$-atom system for realizing a many-body entanglement by the adiabatic tools.

\section{Acknowledgement}

This work is supported by National Key Research Program of China under Grant No. 2016YFA0302000 and by the NSFC under Grants No. 11474094, No. 11104076, No. 11234003, the
Specialized Research Fund for the Doctoral Program of Higher Education No.
20110076120004.

\end{document}